\documentclass[preprint,12pt]{elsarticle}

\usepackage{kurz}
\usepackage{amssymb}
\usepackage{mathrsfs}
\usepackage{multirow}    
\usepackage{stmaryrd}
\usepackage{amsmath}
\usepackage{caption}  
\usepackage{a4wide}
\biboptions{numbers,sort&compress}
\usepackage{upgreek}
\usepackage{longtable}
\usepackage{xcolor}

\usepackage{arydshln}
\usepackage{mathtools}

\journal{Theoretical and Applied Fracture Mechanics}

\begin{document}
\begin{frontmatter}



\title{A micropolar peridynamics model with non-unified horizon for damage of solids with different non-local effects}


\author[Hebei]{Yiming Zhang\corref{cor}}

\author[Hebei]{Xueqing Yang}

\author[tj,tj2]{Xiaoying Zhuang}

\address[Hebei]{School of Civil and Transportation Engineering, Hebei University of Technology, Xiping Road 5340, 300401~Tianjin,~P.R.China }
\address[tj]{Department of Geotechnical Engineering, Tongji University, Siping Road 1239, 200092~Shanghai,~P.R.China}
\address[tj2]{State Key Laboratory for Disaster Reduction in Civil Engineering, Tongji University, Siping Road 1239, 200092~Shanghai,~P.R.China}

\begin{abstract}
Most peridynamics models adopt regular point distribution and unified horizon, limiting their flexibility and engineering applications.  In this work, a micropolar peridynamics approach with non-unified horizon (NHPD) is proposed.  This approach is implemented in a conventional finite element framework, using element-based discretization.  By modifying the dual horizon approach into the pre-processing part, point dependent horizon and non-unified beam-like bonds are built.  By implementing a domain correction strategy, the equivalence of strain energy density is assured.  Then, a novel energy density-based failure criterion is presented which directly bridges the critical stretch to the mechanical strength.  The numerical results indicate the weak mesh dependency of NHPD and the effectiveness of the new failure criterion.  Moreover, it is proven that damage of solid with different non-local effects can lead to similar results by only adjusting the mechanical strength.

\end{abstract}

\end{frontmatter}

\section{Introduction}
\label{sec:intro}
Cracks greatly degrade the durability of structures.  Predicting the initiations and propagations of cracks can assist the researchers to design better structures and maintain their performances.  While partial differential equations are commonly used for describing the balance relations of continuum physical fields, Cracks, i.e. discontinuities will introduce singularities in these equations.  Hence it is a challenge to account in cracks in conventional continuum-based frameworks.  

Despite of the difficulties, in last decades many researchers presented numerical approaches for taking into account discontinuities in a continuous-discontinuous framework, such as remeshing and interface elements \cite{Areias:01,Areias:02,MEJIASANCHEZ2020405}, numerical manifold method and extended finite element methods \cite{ZhengHong:04,YangYongtao:01,Wu2019,SongandBelytschko,WuJianying:01}, cracking elements method \cite{Yiming:11,Yiming:14,Yiming:20}, phase field method \cite{Miehe:01,Wujianying2018a,Wujianying:04}, and particle based methods \cite{Rabczuk:04,Rabczuk:05,Rabczuk20102437}.  Most of these methods deal with discontinuities by moving boundaries, localizing strain, or smearing damaging.  They mostly treat the continuous and discontinuous domains in different manners.

Differently, peridynamics (PD) is a non-local theory solving the continuous-discontinuous problems in the same framework \cite{Silling2000,Hanfei2011}.  It uses the integral balance equations and avoids the singularity of partial differential equations when discontinuities appear.  In the three types of PD formulation: bond-based, ordinary state-based, and non-ordinary state-based, bond-based peridynamics (BBPD) is the first proposed and the most popular one \cite{BBPDreview,Trageser2020}.  After introducing extra parameters or local/global rotational freedom degrees, the fixed Poisson's ratio problem was solved \cite{Zhuqizhi:01,Zhuqizhi:02,Gu2020,Yuhaitao2020}.  Moreover, the numerical procedure of BBPD shows similarity to the other classical lattice element models \cite{Zhaogaofeng2017,Nikolic2018}, easing its implementations.  From numerical point of view, when considering PD as a special type of meshfree method, the bond can be consider as a medium for accelerating the integration process \cite{Bessa2014}.  Despite of the great success of PD, most PD formulations use regular grids of material points and unified horizon, which greatly limits the application and flexibility of PD methods.

In this work, inspired by the dual-horizon peridynamics proposed in \cite{Ren2016,Ren2017}.  The bond-based micropolar peridynamics with shear deformability \cite{DIANA2019201,DIANA2019140} was modified into a peridynamics approach with non-unified horizon (NHPD).  The approach is built in the framework of conventional finite element method (FEM), where the bonds are treated as beam elements.  By using the standard pre-processing step of FEM, it is proven that dual-horizon processing is a part of the modeling (pre-processing) procedure.  The irregular meshes built by Gmsh \cite{Gmsh} are used for providing the peridynamics model with point-dependent horizon, the sizes of which can vary in space greatly.  Then, an iterative domain correction strategy modified from the surface correction strategy is proposed, insuring the equivalence of strain energy density.  Finally, a novel energy density based failure criterion is proposed, correlating the critical stretch to the experimental obtained mechanical strength.  Comparing to the former PD methods, the proposed approach shows great flexibility regarding discretization.  Some benchmark tests are considered, indicating its reliability and robustness.  Last but not least, results in the numerical tests show inspiring correlations among the size of horizon, stiffness, and strength of the structures.  With the proposed framework, the damage of solid with different non-local effects can give similar results.

The remaining parts of this paper are organized as follows: in Section~\ref{sec:method}, the NHPD is presented in details including the pre-processing, beam-like elemental matrix, iterative domain correction strategy, and the damage criterion.  In Section~\ref{sec:ne}, several benchmark tests are used for demonstrating its robustness and reliability.  And the relationship among size of horizon, stiffness and strength is revealed  Finally, Section~\ref{sec:conc} contains concluding remarks.
\section{The micropolar peridynamics with non-unified horizon (NHPD)}
\label{sec:method}
The theories of peridynamics, dual-horizon peridynamics, and correlated non-local operator can be found in such as \cite{Pdtheory,REN2020106235,REN2020113132}, which will not be provided in details in this work.  In this Section, the NHPD will be proposed in a way like conducting a numerical simulation step by step, from modeling to calculation.
\subsection{Pre-precessing and modeling}
\begin{figure}[htbp]
	\centering
	\includegraphics[width=0.7\textwidth]{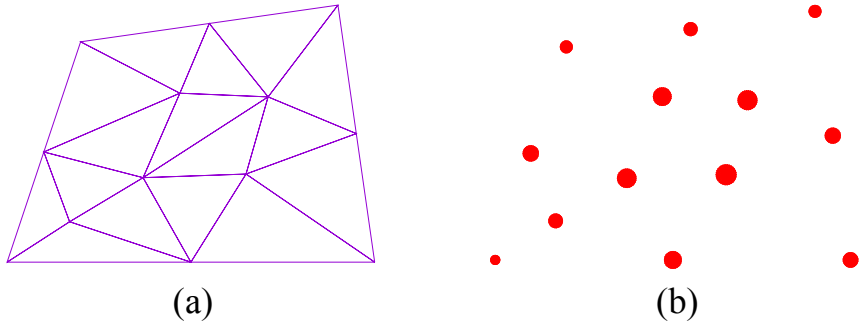}
	\caption{Discretization from elements to material points: (a) planar elements, (b) material points (sizes of the circle indicates their volumes, which are scaled for illustration)}
	\label{fig:FEMMP}
\end{figure}

For pre-precessing, standard FEM discretization is used in this approach.  The domain is discretized into elements, then the nodes are transformed into material points, assuring the equivalence of volume.  For example, in Figure~\ref{fig:FEMMP}, the domain is discretized by linear triangular elements.  Then the volume of every triangular is divided equally to the three nodes, providing the material points.  In this step, other types of elements can also be used, such as Voronoi diagrams \cite{GuXin:01}.  

\begin{figure}[htbp]
	\centering
	\includegraphics[width=0.9\textwidth]{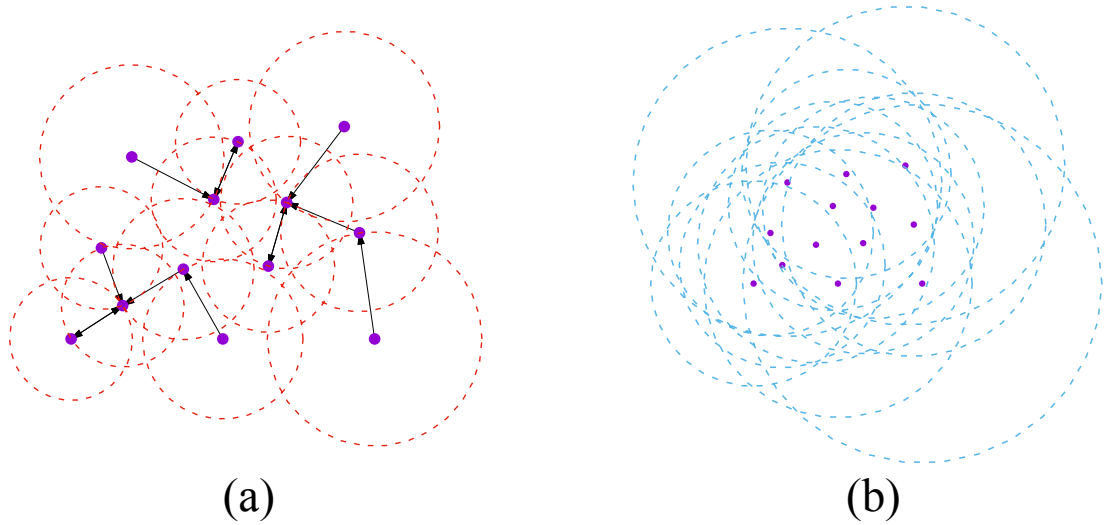}
	\caption{Determination of the horizon: (a) Shortest distances $d_{min}$ between points, (b) Non-unified horizon with $\lambda$=3}
	\label{fig:NHoz}
\end{figure}

Then a point to point distance checking is run for obtaining the distance between the present point to the nearest point, denoted as $d_{min}$.  The example illustrated in Figure~\ref{fig:FEMMP} is used again.  The shortest distances between points are shown in Figure~\ref{fig:NHoz}(a) where the arrows pointed from the present point to its nearest point.  For every material point, the size of its horizon equals to $\lambda \ d_{min}$ and $\lambda\ge 1$ is a prescribed factor.  Regarding $\lambda=3$, the non-unified horizon of the example is illustrated in Figure~\ref{fig:NHoz}(b).  Herein, $\lambda$ can be considered as a non-local parameter in NHPD model.  With the increasing of $\lambda$, the non-local effect of PD model is enhanced.  The influences of $\lambda$ is studied in the numerical studies.

\begin{figure}[htbp]
	\centering
	\includegraphics[width=0.7\textwidth]{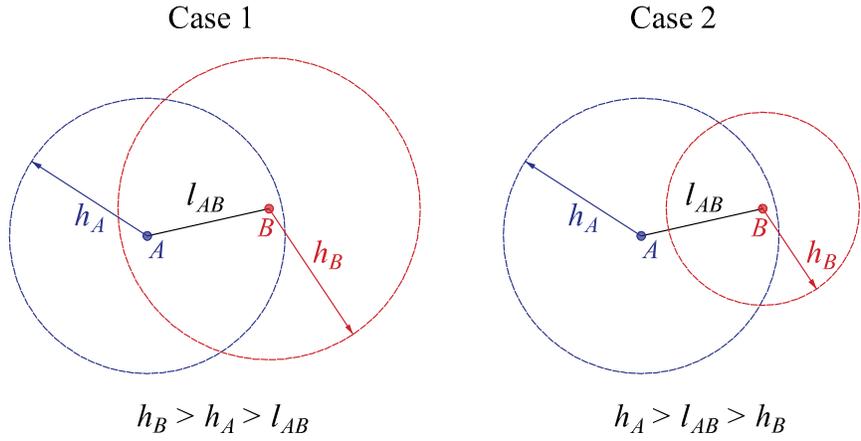}
	\caption{Two cases regarding a bond connecting two points with different sizes of horizon}
	\label{fig:BondHoz}
\end{figure}

After determining the sizes of horizon of every points, the bonds will be built.  When a point locates in the horizon of another point, a bond connecting these two points will be introduced.  The size of horizon will be used in the elemental matrix of the bond.  For the conventional BBPD models, the horizon of bonds equal to the unified horizon.  However in this approach the horizon of bonds are non-unified as well, which are determined locally.  Assuming a bond connects two points: $A$ and $B$ with the sizes of horizon $h_A$ and $h_B$ respectively and the length of this bond is $l_{AB}$.  Then there are two conditions: i) both points locate inside the horizon of the other, ii) one point locates inside the horizon of the other, while the other point locate outside, see Figure~\ref{fig:BondHoz} for example.  The size of the horizon of this bond is 
\begin{equation}
H_{AB}=\left\{\begin{array}{ll}
\left(h_A+h_B\right)\ /\ {2}&\ h_B>h_A>l_{AB},\\
h_A&\ h_A>l_{AB}>h_B.
\end{array}\right.
\label{eq:horizonofbond}
\end{equation}
$H_{AB}$ will be used for building the stiffness matrix of the bond $A-B$.  When building the bonds, the forces are pairwise introduced.  There will be no ghost forces in the domain.  The proposed procedure is consistent with the dual-horizon peridynamics \cite{Ren2016,Ren2017}.  In our model, this procedure is done in the pre-processing step, which is simpler.

\subsection{Beam-like elemental matrix}
The micropolar peridynamics model with shear deformability proposed in \cite{DIANA2019201,DIANA2019140} is used in this approach.  Still, the bond $A-B$ connecting points $A$ and $B$ is considered.  For 2D condition, the coordinations of $A$ and $B$ are $\left[x_A,y_A\right]^T$ and $\left[x_B,y_B\right]^T$.  The rotational matrix is defined as 

\begin{equation}
\mathbf{R}_{AB}=\left[\begin{array}{cccccc}
	a&b&&&\\
	-b&a&&&\\
	 &&1&&&\\
	&&&a&b&\\
	&&&-b&a&\\
	&&&&&1\\
\end{array}\right],
\label{eq:rotR}
\end{equation}
where $a=\left(x_B-x_A\right)\ /\ l_{AB}$ and $b=\left(y_B-y_A\right)\ /\ l_{AB}$.  $x_A$ and $y_A$ are the x and y coordinates of point $A$.  $\mathbf{R}_{AB}$ is the same as the rotational matrix used in the beam element.  

Three degrees of freedom: two displacements along two axis and one rotational displacement are considered on each point, denoted as $\mathbf{u}_A=\left[ux_{A},uy_{A},m_A\right]^T$ and $\mathbf{u}_B=\left[ux_{B},uy_{B},m_B\right]^T$ respectively.  By mimicking the beam element, the normal, shearing, and rotational deformations of the bond are denotes as $\left[s, \gamma, \vartheta\right]^T$.  For bond $A-B$, 
\begin{equation}
\begin{aligned}
&\left[\begin{array}{c}
s\\
\gamma\\
\vartheta
\end{array}\right]_{AB}=\mathbf{B}^T_{AB} \ \mathbf{R}_{AB} \ 
\left[\begin{array}{c}
\mathbf{u}_A\\
\mathbf{u}_B
\end{array}\right],\\
&\mbox{where}\\
&\mathbf{B}^T_{AB}=\frac{1}{l_{AB}}\left[\begin{array}{cccccc}
-1&0&0&1&0&0\\
0&-1&l_{AB}\ /\ 2&0&1&l_{AB}\ /\ 2\\
0&0&-l_{AB}&0&0&l_{AB}
\end{array}\right].\\
\label{eq:bonddeform}
\end{aligned}
\end{equation}

Then, during elastic loading the normal, shearing, and momentum forces of the bond are denoted as $\left[f_n, f_t, m_\vartheta\right]^T$.  For bond $A-B$,

\begin{equation}
\left[\begin{array}{c}
f_n\\
f_t\\
m_\vartheta
\end{array}\right]_{AB}=\Omega_{AB} \ \alpha_{AB} \ \mathbf{D}_{AB} \ 
\left[\begin{array}{c}
s\\
\gamma\\
\vartheta
\end{array}\right]_{AB},\\
\label{eq:bondstress}
\end{equation}
where $\mathbf{D}_{AB}$ is the spring equivalent matrix as
\begin{equation}
\mathbf{D}_{AB}=\left[\begin{array}{ccc}
k_{n,AB}&&\\
&k_{t,AB}&\\
&&k_{\vartheta,AB}
\end{array}\right].
\label{eq:springD}
\end{equation}
In Eq.~\ref{eq:bondstress}, $k_{n,AB}$, $k_{t,AB}$, and $k_{\vartheta,AB}$ are the bond normal, shearing, and rotational spring equivalent stiffness factors, determined by
\begin{equation}
\begin{aligned}
&k_{n,AB}=c_{AB}\\
&k_{t,AB}={12\ d_{AB}}\ / \ {l_{AB}^2}\\
&k_{\vartheta,AB}={d_{AB}}\ / \ {l_{AB}}\\
&\mbox{where}\\
&c_{AB}=\left\{\begin{array}{ll}
\cfrac{6\ E}{\pi\ t \ H_{AB}^3\ \left(1-\nu\right)}&\mbox{   plane stress}\\
&\\
\cfrac{6\ E}{\pi\ t \ H_{AB}^3\ \left(1-2\ \nu\right)\ \left(1+\nu\right)}&\mbox{   plane strain}
\end{array}
\right.,\\
&\mbox{and}\\
&d_{AB}=\left\{\begin{array}{ll}
\cfrac{E\ \left(1-3\ \nu\right)}{6\ \pi\ t \ H_{AB}\ \left(1-\nu^2\right)}&\mbox{   plane stress}\\
&\\
\cfrac{E\ \left(1-4\ \nu\right)}{6\ \pi\ t \ H_{AB}\ \left(1-2\ \nu\right)\ \left(1+\nu\right)}&\mbox{   plane strain}
\end{array}
\right.,
\label{eq:kfactors}
\end{aligned}
\end{equation}
where $E$ is the elastic modulus and $\nu$ is the Poisson's ratio.  $t$ is the thickness.  For simplicity, $t=1$~m is considered in this work.  

In Eq.\ref{eq:bondstress}, $\alpha_{AB}$ is the length correction coefficient.  It is introduced for accounting the influences of bonds with different lengths.  The short bonds are considered to have greater influences on the mechanical responses than the long bonds.  $\alpha_{AB}$ is determined by taking the mean value of the normalized values of $l_{AB}$ regarding points $A$ and $B$ as
\begin{equation}
\alpha_{AB}=\frac{1}{2}\left[ \mbox{exp}\left(\frac{l_{Amin}-l_{AB}}{l_{Amax}-l_{Amin}}\right)+ \mbox{exp}\left(\frac{l_{Bmin}-l_{AB}}{l_{Bmax}-l_{Bmin}}\right)\right],
\end{equation}
where $l_{Amax}$ and $l_{Amin}$ are the maximum and minimum lengths of bonds connecting to point $A$, and $l_{Bmax}$ and $l_{Bmin}$ are the maximum and minimum lengths of bonds connecting to point $B$.  $\Omega_{AB}$ is the domain correction coefficient which will be discussed in the next section.

Correspondingly, the potential energy for
the bond $A-B$, denoted as $E_{AB}$, is determined by
\begin{equation}
E_{AB}=\int\int \frac{f_n\ s\ l_{AB}}{2}+\frac{f_t\ \gamma\ l_{AB}}{2}+\frac{m_\vartheta\ \vartheta}{2}\ dV_A \ dV_B,
\label{eq:EAB}
\end{equation}
in which, $l_{AB}$ appears only in the first two terms.  Then, the beam-like elemental stiffness matrix $\mathbf{K}_{AB}$ is determined by
\begin{equation}
\begin{aligned}
&\mathbf{K}_{AB}=\Omega_{AB} \ \alpha_{AB} \ V_{A} \ V_{B} \ \mathbf{R}_{AB}^T\ \mathbf{B}_{AB}\ \mathbf{L}_{AB}\ \mathbf{D}_{AB} \ \mathbf{B}_{AB}^T\ \mathbf{R}_{AB},\\
&\mbox{where}\\
&\mathbf{L}_{AB}=\left[\begin{array}{ccc}
l_{AB}&&\\
&l_{AB}&\\
&&1
\end{array}\right].
\label{eq:stiffnessK}
\end{aligned}
\end{equation}
In Eq.~\ref{eq:stiffnessK}, $V_{A}$ and $V_{B}$ are the volumes of the material points $A$ and $B$ respectively.  $\mathbf{K}_{AB}$ will be assembled into the global stiffness matrix one after another, just like the conventional FEM models.

Correspondingly $E_{AB}$ can be approximately determined by
\begin{equation}
E_{AB}=\frac{1}{2}\left(\left[\mathbf{u}_A\ \mathbf{u}_B\right]\mathbf{K}_{AB}\left[\begin{array}{c}
\mathbf{u}_A\\
\mathbf{u}_B
\end{array}\right]\right).
\end{equation}

\subsection{Iterative domain correction strategy}
The iterative domain correction strategy is inspired by the energy-based surface correction strategy \cite{Pdtheory,Quang2019}, which is used for correcting the stiffness of the surface material points whose non-local effects are different from those of the inner points.  When non-unified horizon inevitably introduce non-homogenized material points and more complex point to point bonds, the equivalence of the strain energy density cannot be insured automatically.  Hence, all bonds in the domain need to be corrected.  Comparing to the original energy-based correction method, another main difference of the proposed strategy is that the correction strategy will be run iteratively.  

Firstly, assuming a domain experiences unified normal strain $\varepsilon$ along a specified direction, the strain energy density $e$ in the domain can be determined by
\begin{equation}
e=\left\{\begin{array}{ll}
\cfrac{E\ \varepsilon^2}{2\ \left(1-\nu^2\right)}&\mbox{   plane stress}\\
&\\
\cfrac{E\ \left(1-\nu\right)\ \varepsilon^2}{2\ \left(1+\nu\right)\left(1-2\ \nu\right)}&\mbox{   plane strain}
\end{array}
\right.,
\label{eq:sed}
\end{equation}
which shall be the true value for each material point.

Furthermore, the trail value of the strain energy density on material point $A$, denoted by $\tilde{e}_{A}$, can be determined by
\begin{equation}
\begin{aligned}
&\tilde{e}_{A}=\frac{1}{4}\sum\limits_{B}^{n}\left\{\left[\mathbf{u}_A\ \mathbf{u}_B\right]\ \frac{\mathbf{K}_{AB,trail}}{V_A}\ \left[\begin{array}{c}
\mathbf{u}_A\\
\mathbf{u}_B
\end{array}\right]\right\}\\
&\mbox{where}\\
&\mathbf{K}_{AB,trail}= \Omega_{AB,j} \ \alpha_{AB}\ V_{B}\ \mathbf{R}_{AB}^T\ \mathbf{B}_{AB}\ \mathbf{L}_{AB}\  \mathbf{D}_{AB} \ \mathbf{B}_{AB}^T\ \mathbf{R}_{AB},
\end{aligned}
\label{eq:etrailA}
\end{equation}
where $1\ /\ 4$ appears because the energy of a bond is shared by two points.  $n$ is the set of all points connecting to point $A$ by bonds.  $\Omega_{AB,j}$ is the value of $\Omega_{AB}$ at correction iteration step $j$ with $\Omega_{AB,0}=1$.  Here we would like to mention that the correction iteration step $j$ does not relate to the Newton-Raphson iteration step.  The iterative domain correction strategy will be conducted before the main calculation starts.  Once $\Omega_{AB}$ is obtained, the values will not change during the calculation.

Then, with Eq.~\ref{eq:etrailA}, firstly applying $\varepsilon$ along the x direction by setting $u_x=\varepsilon\ x$, then applying $\varepsilon$ along the y direction by setting $u_y=\varepsilon\ y$, correspondingly the strain energy density of points $A$ and $B$ along x and y direction: $\tilde{e}_{A,x}$, $\tilde{e}_{A,y}$ and $\tilde{e}_{B,x}$, $\tilde{e}_{B,y}$ will be obtained.  With these values, the domain correction factor of bond $A-B$, $\Omega_{AB}$ is determined by
\begin{equation}
\begin{aligned}
&\Omega_{AB,j+1}=\frac{\Omega_{AB,j}}{\sqrt{\left({a}\ / \ {p_x}\right)^2+\left({b}\ / \ {p_y}\right)^2}}\\
&\mbox{with}\\
&p_x=\frac{1}{2}\left(\frac{\varepsilon}{\tilde{e}_{A,x}}+\frac{\varepsilon}{\tilde{e}_{B,x}}\right),\ p_y=\frac{1}{2}\left(\frac{\varepsilon}{\tilde{e}_{A,y}}+\frac{\varepsilon}{\tilde{e}_{B,y}}\right),
\end{aligned}
\label{eq:omegaab}
\end{equation}
where $a$ and $b$ are the same as denoted in Eq.~\ref{eq:rotR}.  When $\sum\limits_{AB}\left\{ |\Omega_{AB,j}-\Omega_{AB,j-1}| \right\}<10^{-3}$, the domain correction strategy will be stopped and $\Omega_{AB}=\Omega_{AB,j}$.

\subsection{Damage model and the implicit iteration}
The peridynamics theory shows differences from the conventional continuum-based method the investigations of which are still undergoing.  There are many different damage criteria on the market, see \cite{YANG2020105830,zaccariotto2015,Foster2011,Huang2015,Zhang2019,DIPASQUALE2017378,RABCZUK201742} for example. 

In this work, a novel energy density based criterion is proposed.  Based on Eq.~\ref{eq:sed}, the critical strain energy density under uni-axial tension with tensile stress equals to the tensile strength $F_t$ is 
\begin{equation}
e_0=\left\{\begin{array}{ll}
\cfrac{F_t^2}{2\ E\  \left(1-\nu^2\right)}&\mbox{   plane stress}\\
&\\
\cfrac{F_t^2\ \left(1-\nu^2\right)\ \left(1-\nu\right)^2}{2\ E\ \left(1-2\ \nu\right)}&\mbox{   plane strain}
\end{array}
\right..
\label{eq:e0}
\end{equation}

\begin{figure}[htbp]
	\centering
	\includegraphics[width=0.95\textwidth]{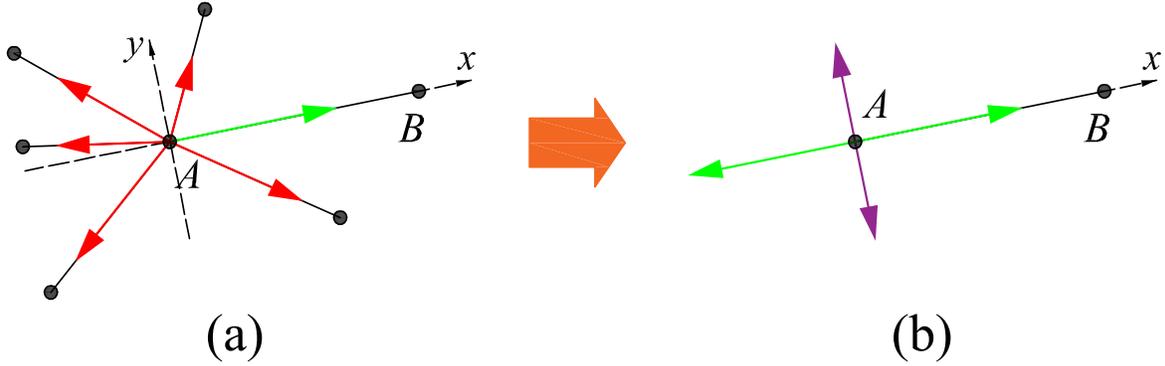}
	\caption{Bonds connecting to point $A$: (a) forces balance at point $A$, (b) decomposed forces along the $x$ and $y$ axes}
	\label{fig:Engeqv}
\end{figure}

Focusing on the point $A$ with bond $A-B$ connecting to another point $B$, the balance relations of the forces of all bonds connecting to $A$ are fulfilled, see Figure~\ref{fig:Engeqv}.  When considering only the stretch $s$ of the bond $A-B$, based on Eq.~\ref{eq:etrailA}, the dedication of the bond $A-B$ to the strain energy density at uni-axial loading condition along the $x$-axis can be considered as half of the total strain energy density $e_A$ as
\begin{equation}
\frac{e_{A}}{2}=\frac{\Omega_{AB}\ \alpha_{AB}\ V_A\ V_B\ c_{AB}\ l_{AB}\ s^2}{2\left(V_A+V_B\right)},
\label{eq:bondeab}
\end{equation}
where the stain energy density at point $B$ is also considered.

Setting $e_{A}=e_0$, the corresponding stretch $s$ is considered to be the critical stretch $s_0$.  When using the same procedure on point $B$, the same result will be obtained.  Finally the critical stretch $s_0$ of the bond $A-B$ is obtained as
\begin{equation}
s_0=\ \sqrt{\frac{e_0\ \left(V_A+V_B\right)}{\Omega_{AB}\ \alpha_{AB}\ V_A\ V_B\ c_{AB}\ l_{AB}}}\ ,
\label{eq:s0value}
\end{equation}
where, the subscript $AB$ is ignored for $s_0$ for simplicity.  Eq.~\ref{eq:s0value} directly correlates the critical stretch to the experimentally obtained mechanical strength, bringing great flexibility for engineering practices.

After determining the critical stretch $s_0$, the isotropic damage model presented for some other PD formulations can also be used, such as the bilinear softening model \cite{DIANA2020106985,XuChen2020} and the exponential softening model \cite{TONG2020106767}.  However, this is beyond the topic of this work.  Hence, conventional prototype microelastic brittle (PMB) is used here that once $s_{AB}\ge s_0$, the bond $A-B$ is assumed to break completely and its damage degree $d_{AB}$ is set to 1 otherwise $d_{AB}=0$.  Hence, for a material point $A$ with some damaged bonds, its point damage degree $\varphi_A$ is defined as
\begin{equation}
\varphi_A=1-\frac{\sum\limits_{B}^{n}\left\{\left(1-d_{AB}\ \right)\ \Omega_{AB}\ \alpha_{AB}\right\}}{\sum\limits_{B}^{n}\left\{\Omega_{AB}\ \alpha_{AB}\right\}}.
\label{eq:dmgA}
\end{equation}
$\varphi$ is determined in the end of every load step, which can be considered as a post-processing step.

\begin{figure}[htbp]
	\centering
	\includegraphics[width=0.75\textwidth]{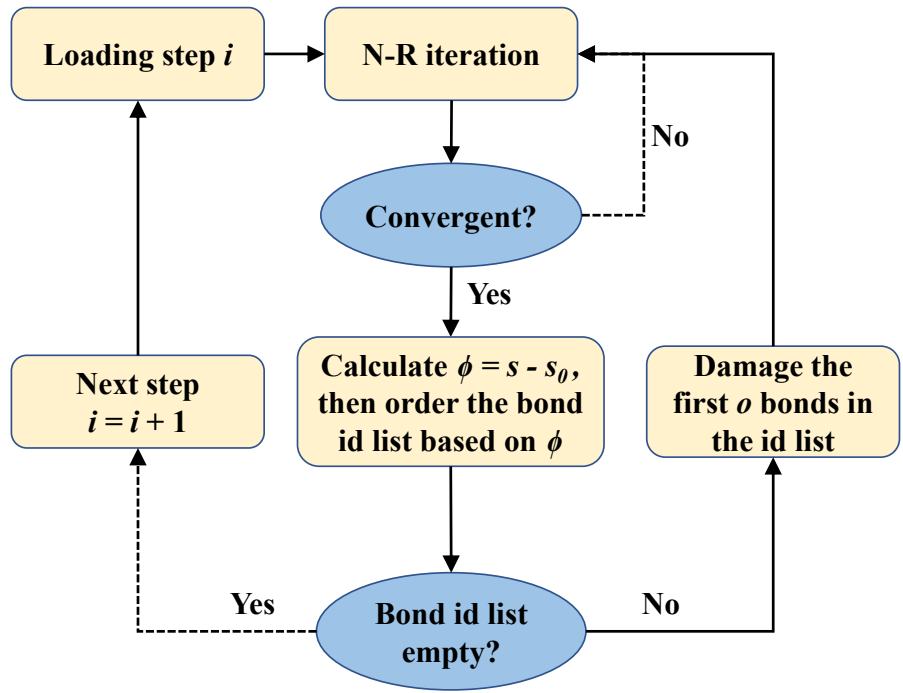}
	\caption{Calculation procedure within one N-R iteration step}
	\label{fig:procedure}
\end{figure}

During the numerical iteration, in one step, damaging too many bonds may result in numerical instability and overestimation of the damage zone.  The implicit iteration procedure is adopted for enhancing the numerical stability \cite{Bie2020}.  For convenience, the following global matrix and vectors are defined:
\begin{equation}
\boldsymbol{\mathsf{K}}=\bigcup \left(\mathbf{K_{AB}}\right)\ \mbox{ and }\ \boldsymbol{\mathsf{U}}=\bigcup \left(\left[\begin{array}{c}
\mathbf{u}_A\\
\mathbf{u}_B
\end{array}\right]\right),
\label{eq:vectorg}
\end{equation}
where $\bigcup\left(\cdot\right)$ denotes the assemblage of the beam-like elemental matrix or vector to the global form.  According to the Newton-Raphson (N-R) method, for the iteration step $j$ at the load step $i$, the element-related incremental relation is 
\begin{equation}
\begin{array}{cccc}
\boldsymbol{\mathsf{U}}_{i,j}=
&\underbrace{\boldsymbol{\mathsf{U}}_{i-1}+\Delta\boldsymbol{\mathsf{U}}_{j-1}}&+&\underbrace{\Delta\Delta\boldsymbol{\mathsf{U}}}\\
&\mbox{known}&&\mbox{unknown}
\end{array},
\label{eq:UdU2}
\end{equation}
in which $\Delta\left(\cdot\right)$ denotes an increment of the corresponding value at the preceding load step, $i-1$, while $\Delta\Delta\left(\cdot\right)$ stands for an increment of the value at the last N-R iteration step, $j-1$.  The same framework is used for building some other types of numerical tools, see \cite{Yiming:15,Yiming:16} for example.  At every iteration step, the balance equation is 
\begin{equation}
\boldsymbol{\mathsf{K}}_j\ \Delta\Delta\boldsymbol{\mathsf{U}}=\boldsymbol{\mathsf{F}}_{i}-\boldsymbol{\mathsf{K}}_j\left(
\boldsymbol{\mathsf{U}}_{i-1}+\Delta\boldsymbol{\mathsf{U}}_{j-1} \right)
\label{eq:balance}
\end{equation}
where $\boldsymbol{\mathsf{F}}_{i}$ is the loading forces at load step $i$.  The total elastic energy of the system $E=\sum\limits_{AB}\left\{E_{AB}\right\}$ is used for checking whether the equilibrium iteration by means of the N-R method converges.  Thus, if
\begin{equation}
\mbox{if } \left|\frac{E_j- E_{j-1}}{E_j} \right|<\epsilon,
\label{eq:converg}
\end{equation}
then the N-R iteration converged at step $j$, where $\epsilon$ is a prescribed small value with $\epsilon=10^{-4}$ in all numerical examples.  When the equilibrium iteration converges, the breakage of bonds will be checked.  The value $\phi=s-s_0$ of every bond is obtained.  Then, the indexes of unbroken bonds with $\phi>0$ are ordered into a list from the biggest to the smallest values of $\phi$.  With a prescribed number $o$, the first $o$ bonds in this list will be broken.  And the N-R iteration will be rerun.  When this list becomes empty in one iteration step, the N-R iteration of this load step converges.  The algorithm of the described procedure is illustrated in Figure~\ref{fig:procedure}.  The computing efficiency will be enhanced with the increasing of $o$ while the numerical stability will be reduced.  $o \le 10$ is recommended.  It can be found that this procedure is similar to that of the cracking elements method \cite{Yiming:20,Yiming:21}, which cracks the element one after another.

\section{Numerical investigations}
\label{sec:ne}
Plane stress condition is considered for all the numerical examples provided in this section.
\subsection{Intact Brazilian disk tests}
\begin{figure}[htbp]
	\centering
	\includegraphics[width=0.95\textwidth]{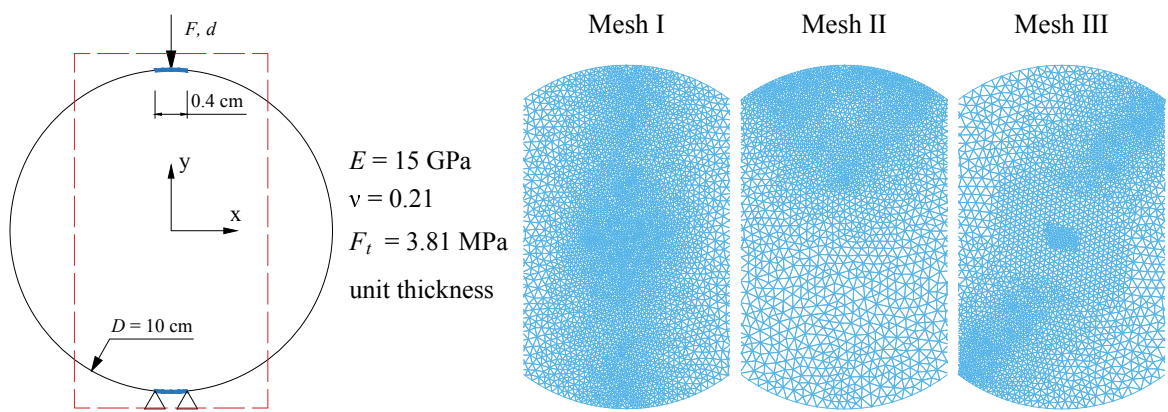}
	\caption{Intact disk test: model, material and meshes}
	\label{fig:DiskIntactModel}
\end{figure}

The model, material and meshes of the intact disk test are shown in Figure~\ref{fig:DiskIntactModel}.  Three meshes are considered.  The analytical peak load per unit thickness is $F_{peak}=\left(\pi \ D\ F_t\right)\ / \ {2}=$ 598.47~\mbox{kN}.  Different values of non-local parameter $\lambda$ are considered.

\begin{figure}[htbp]
	\centering
	\includegraphics[width=0.9\textwidth]{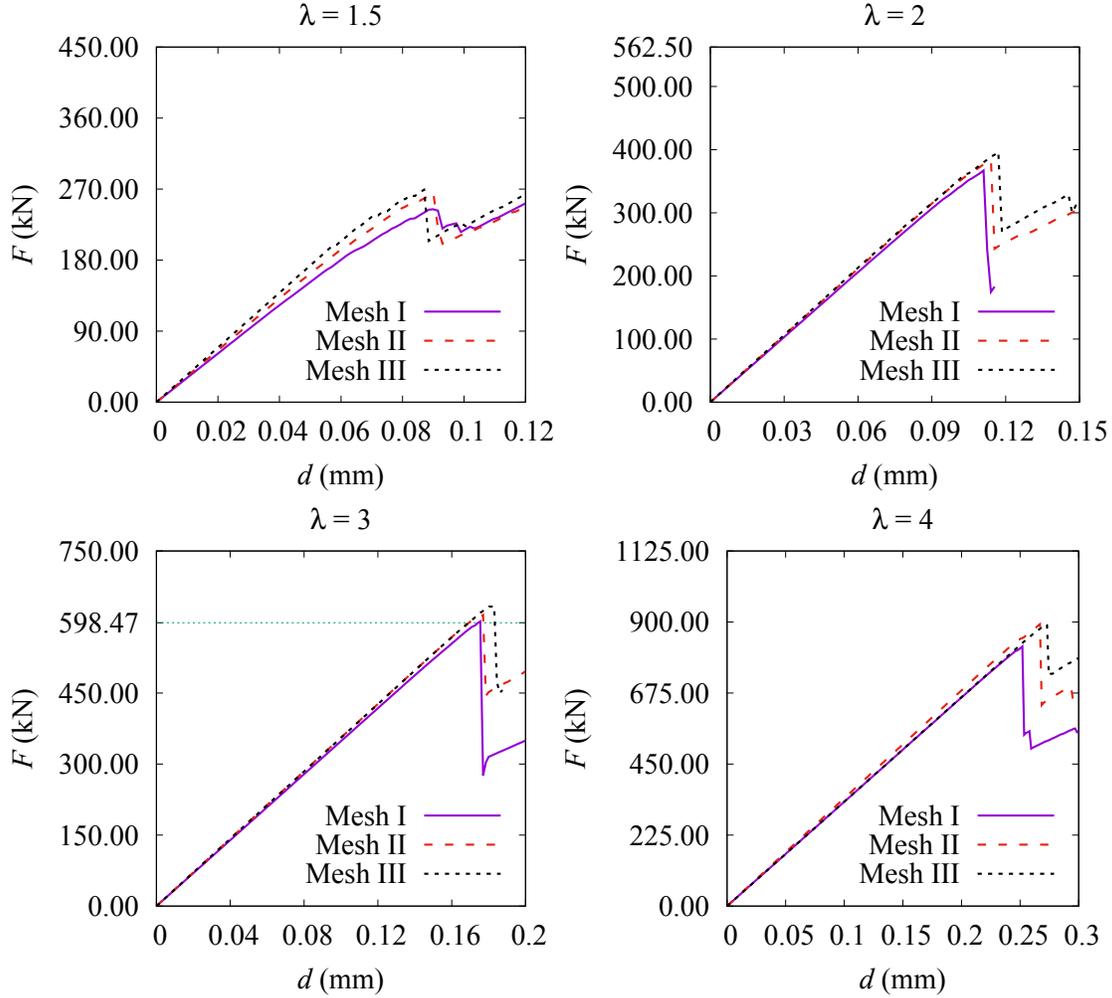}
	\caption{Intact disk test: force-displacement curves considering $\lambda=1.5$, $\lambda=2$, $\lambda=3$, and $\lambda=4$}
	\label{fig:DiskIntactForce}
\end{figure}

\begin{figure}[htbp]
	\centering
	\includegraphics[width=0.5\textwidth]{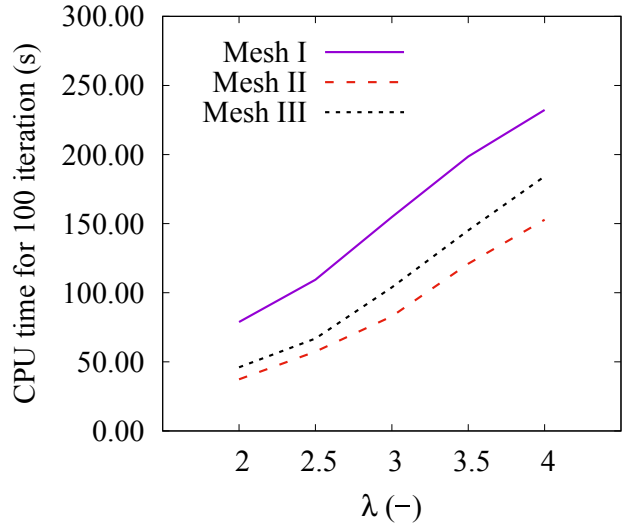}
	\caption{Relationship between computing time and $\lambda$}
	\label{fig:CPUtime}
\end{figure}

The force-displacement curves are shown in Figure~\ref{fig:DiskIntactForce}.  From the results it can be found:
\begin{itemize}
	\item 
	When $\lambda\ge 2$, the stiffness of the structure are generally similar with different $\lambda$;
	\item 
	When $\lambda\ge 2$, the stiffness of the structure are generally similar with different meshes;
	\item
	The values of the peak load increase considerably with the increasing of $\lambda$;
	\item
	The values of the peak load are slightly different with different meshes;
	\item
	When $\lambda=3$, the values of the peak load approach the analytical value.
\end{itemize}
The final finding coincides with the common assumption used in most PD model using unified grid points that the horizon should be around three times of the spacing of the material points.  On the other hand, for specific mesh, though the order of the global stiffness matrix is the same, the computing time will increase with the increasing of $\lambda$, see Figure~\ref{fig:CPUtime}.  Because when $\lambda$ increases, the global stiffness matrix becomes denser.

\begin{figure}[htbp]
	\centering
	\includegraphics[width=0.6\textwidth]{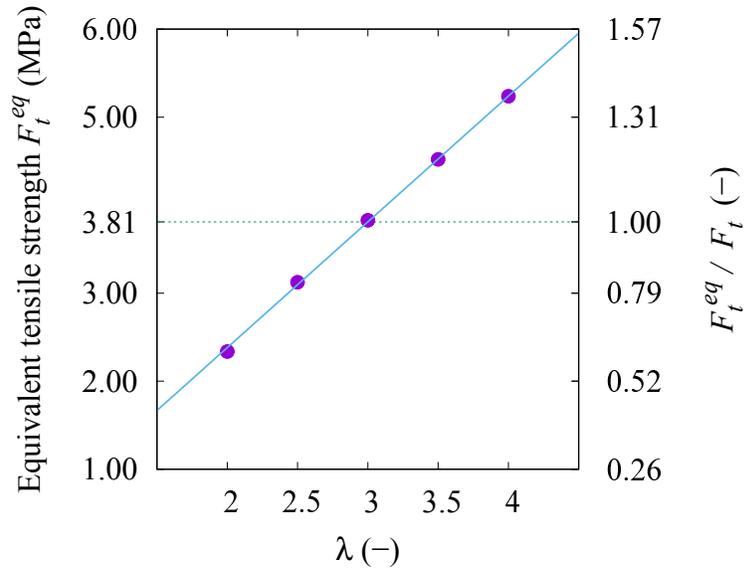}
	\caption{Relationship between equivalent tensile strength $F_t^{eq}$ and $\lambda$ (results of Mesh I)}
	\label{fig:Lft}
\end{figure}

Furthermore, we follow the finding that the peak load changes with $\lambda$.  Considering the results with Mesh I, we obtain the equivalent tensile strength $F_t^{eq}$ from $F_{peak}$, depending on $\lambda$.  The results is illustrated in Figure~\ref{fig:Lft} where the fitting curve is 
\begin{equation}
\frac{F_t^{eq}}{F_t}=\frac{3\ \lambda-1}{8}.
\label{eq:Lft}
\end{equation}

\begin{figure}[htbp]
	\centering
	\includegraphics[width=0.9\textwidth]{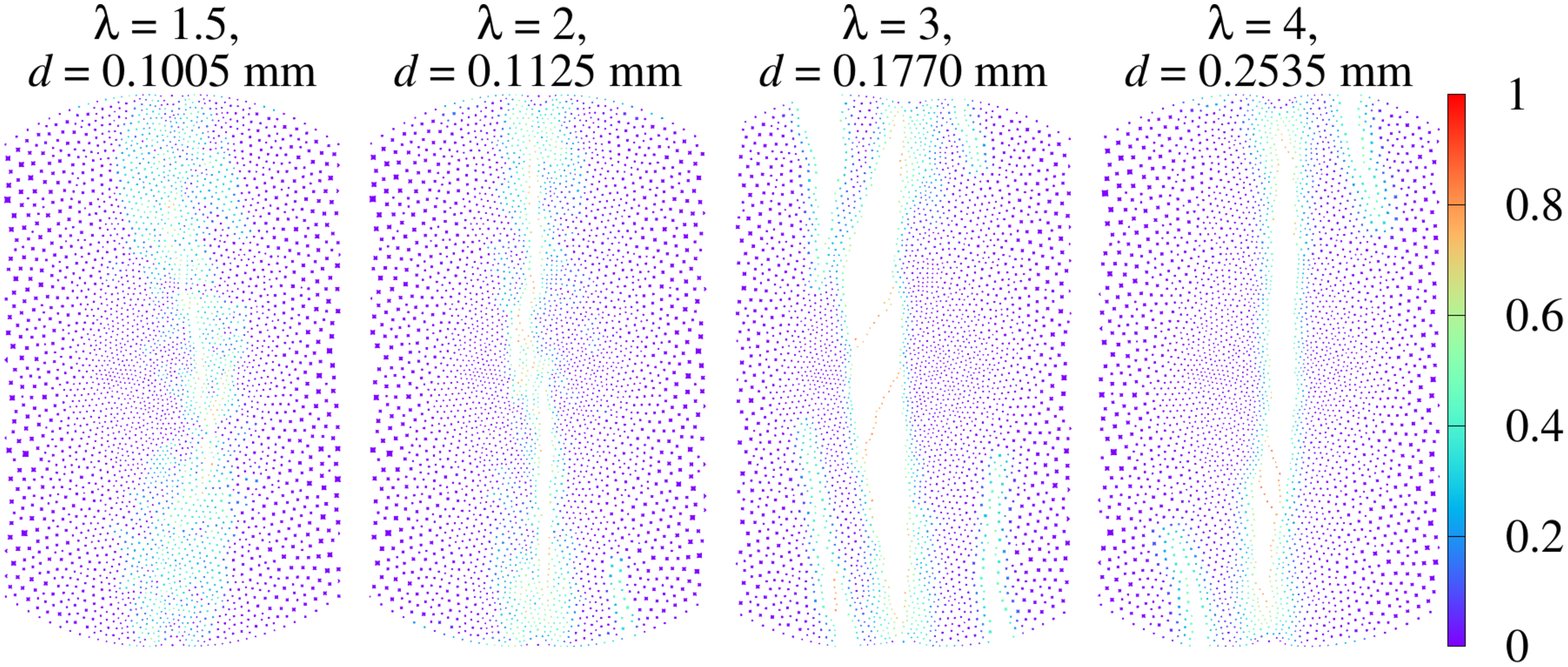}
	\caption{Intact disk test: damage degree and deformation plots (scale: 1:20) of Mesh I}
	\label{fig:DiskIntactDmgMeshI}
\end{figure}

\begin{figure}[htbp]
	\centering
	\includegraphics[width=0.9\textwidth]{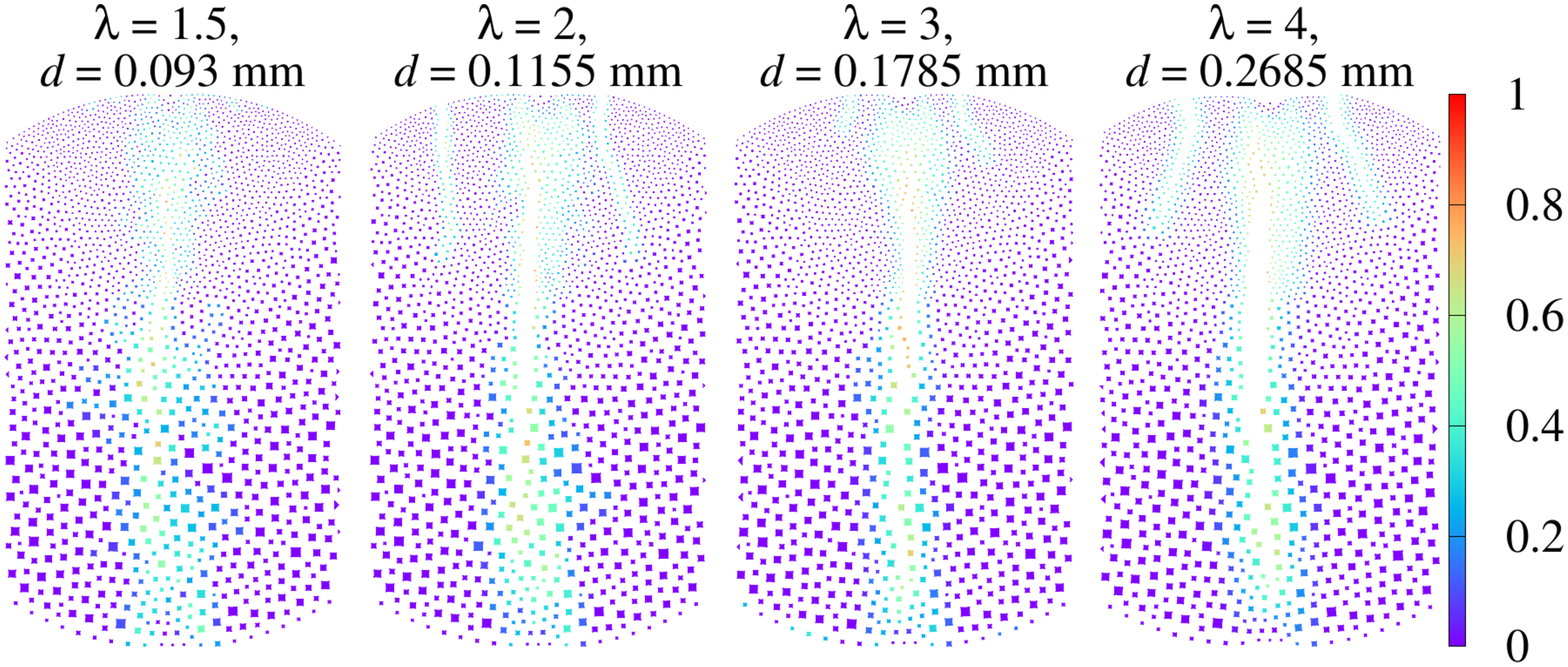}
	\caption{Intact disk test: damage degree and deformation plots (scale: 1:20) of Mesh II}
	\label{fig:DiskIntactDmgMeshII}
\end{figure}

\begin{figure}[htbp]
	\centering
	\includegraphics[width=0.9\textwidth]{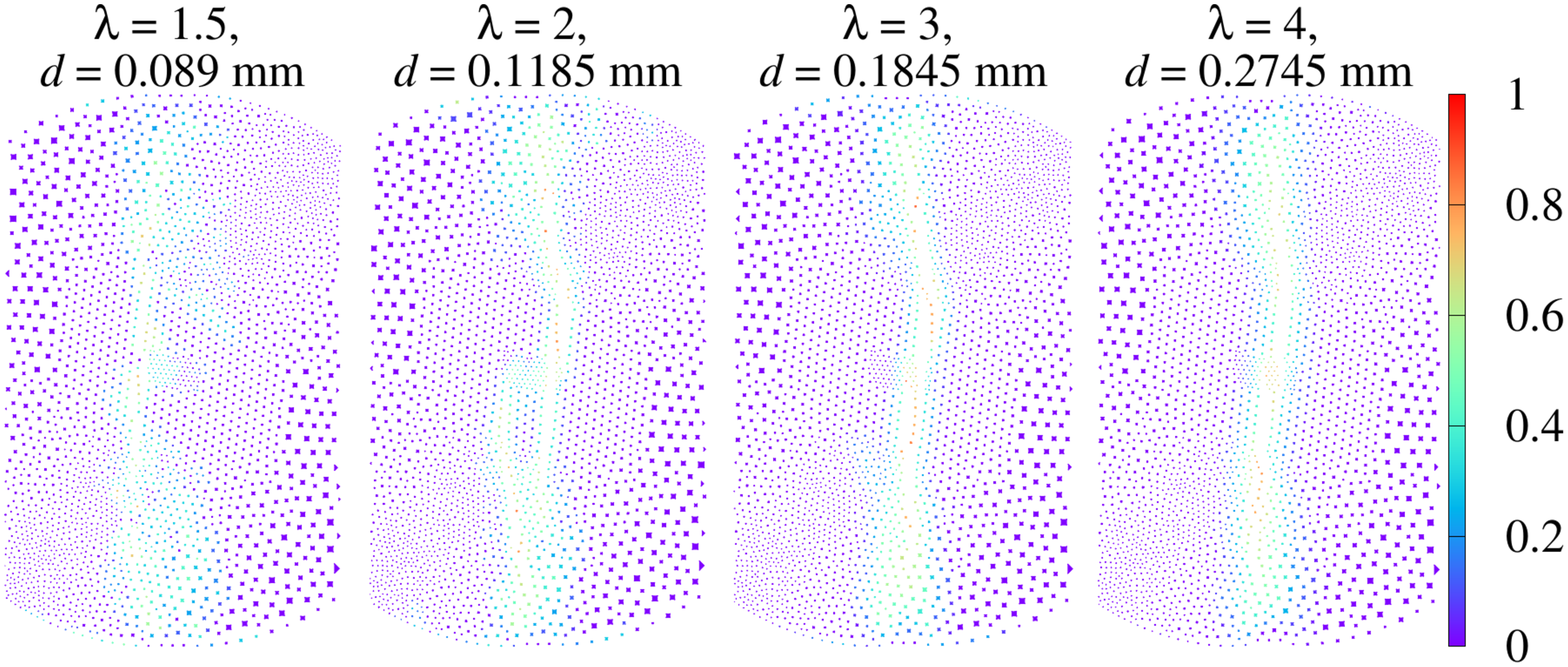}
	\caption{Intact disk test: damage degree and deformation plots (scale: 1:20) of Mesh III}
	\label{fig:DiskIntactDmgMeshIII}
\end{figure}

Finally, the damage degree and deformation plots are shown in Figures~\ref{fig:DiskIntactDmgMeshI} to~\ref{fig:DiskIntactDmgMeshIII}.  Generally similar and reasonable failure patterns are found.  The damage initiate from the middle of the disk, but not from the boundary.  This is agreeable with the experiments.  Moreover, it proves that the boundary weakness of PD model is greatly mitigated in the NHPD.

\subsection{Brazilian disk tests with slots}
\begin{figure}[htbp]
	\centering
	\includegraphics[width=0.85\textwidth]{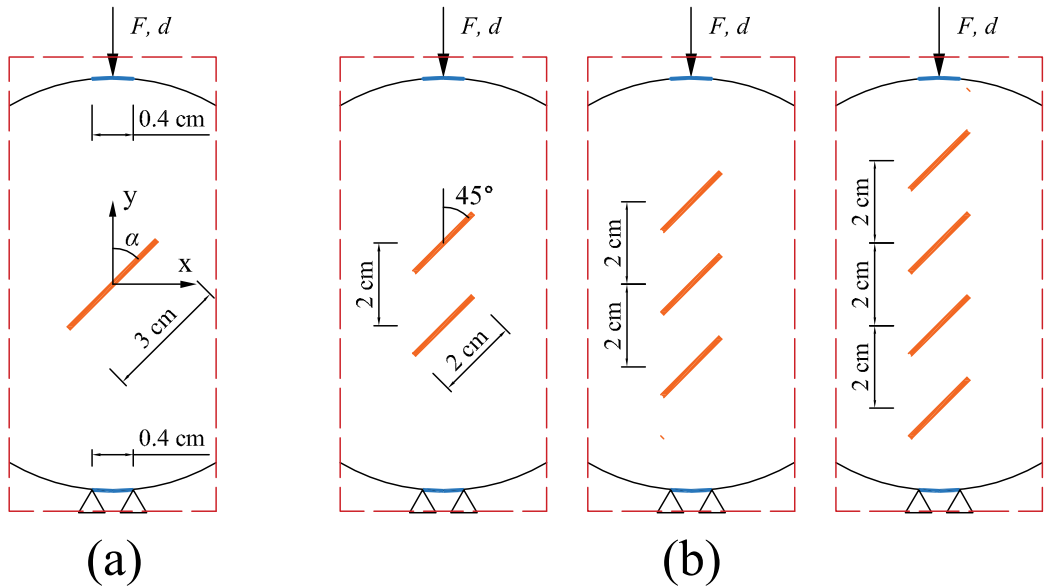}
	\caption{Disk tests with slots: model (a) disk tests with a single slot (different angles of inclination $\alpha$), (b) disk tests with multiple slots}
	\label{fig:DiskSlotsModel}
\end{figure}

Brazilian disk tests with a single slot and multiple slots were experimentally investigated in \cite{HAERI201420,Haeri2015}.  The models are shown in Figure~\ref{fig:DiskSlotsModel}.  Mesh I shown in Figure~\ref{fig:DiskIntactModel} is used.  The slots are not explicitly modeled but the bonds intersect with the slots are removed.  About the material properties, same values of $E$, $\nu$, and $\eta$ as taken in the last example are used.  On the other hand, we consider different values of $\lambda$.  Hence the values of $F_t$ is adjusted based on Eq.~\ref{eq:Lft}, as: i) $\lambda=2.5, F_t=4.69$~MPa, ii) $\lambda=3, F_t=3.81$~MPa, and iii) $\lambda=3.5, F_t=3.21$~MPa.  $598.47~\mbox{kN}$ is used for obtaining normalized peak loads for all cases.  

\begin{figure}[htbp]
	\centering
	\includegraphics[width=0.85\textwidth]{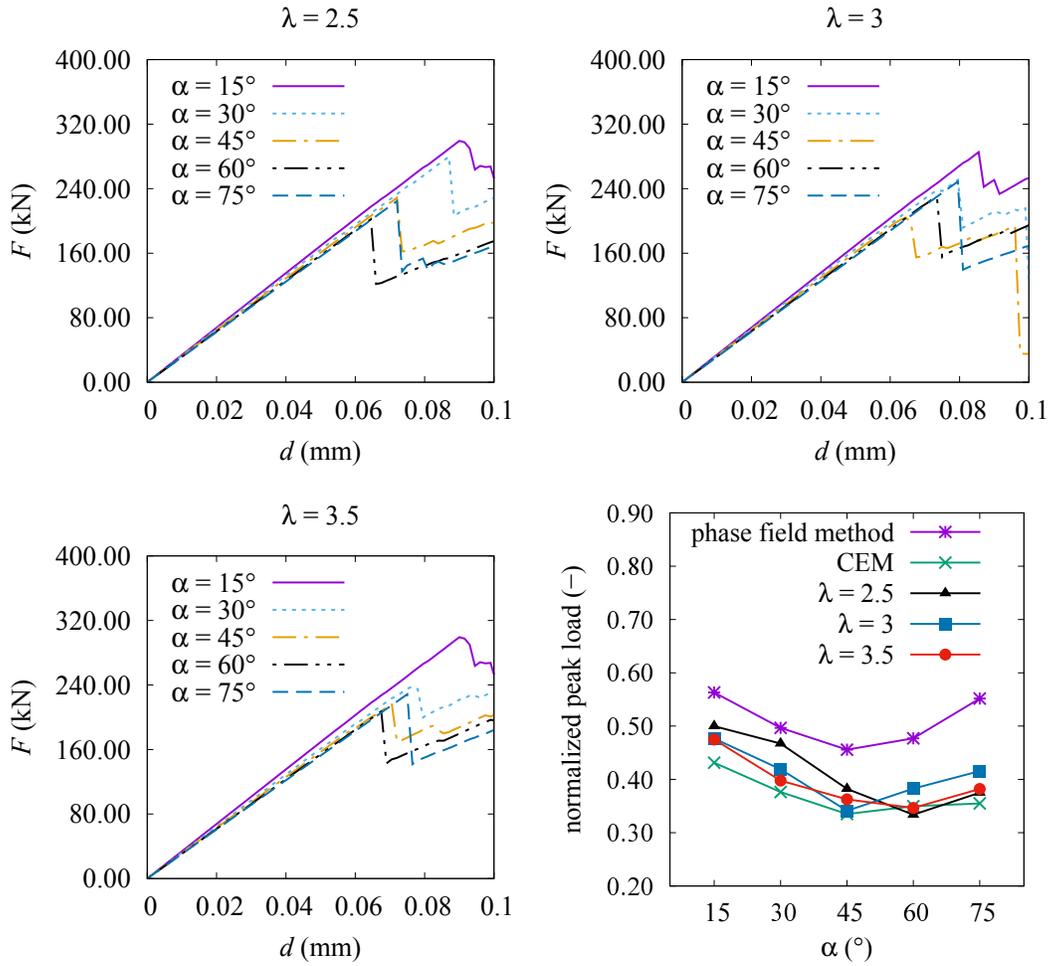}
	\caption{Disk tests with an inclined slot: force-displacement curves and normalized peak loads comparing to the results obtained with phase field method provided in \cite{Zhou2019} and CEM in \cite{Yiming:20}}
	\label{fig:Disk1slotForces}
\end{figure}

\begin{figure}[htbp]
	\centering
	\includegraphics[width=0.75\textwidth]{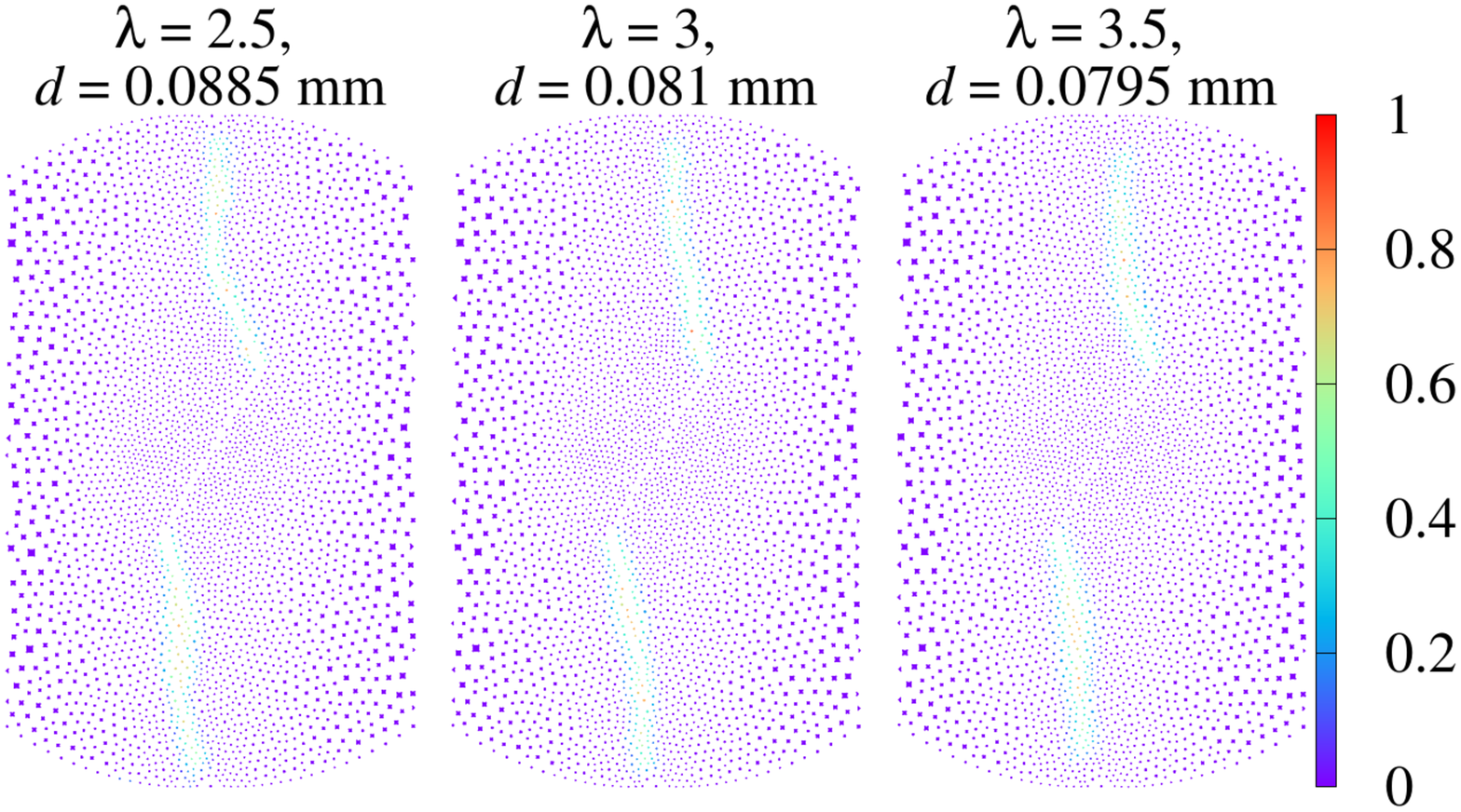}
	\caption{Disk tests with an inclined slot: damage degree and deformation plots (scale: 1:20) with $\alpha=30^\circ$}
	\label{fig:Disk1slot30deg}
\end{figure}

\begin{figure}[htbp]
	\centering
	\includegraphics[width=0.75\textwidth]{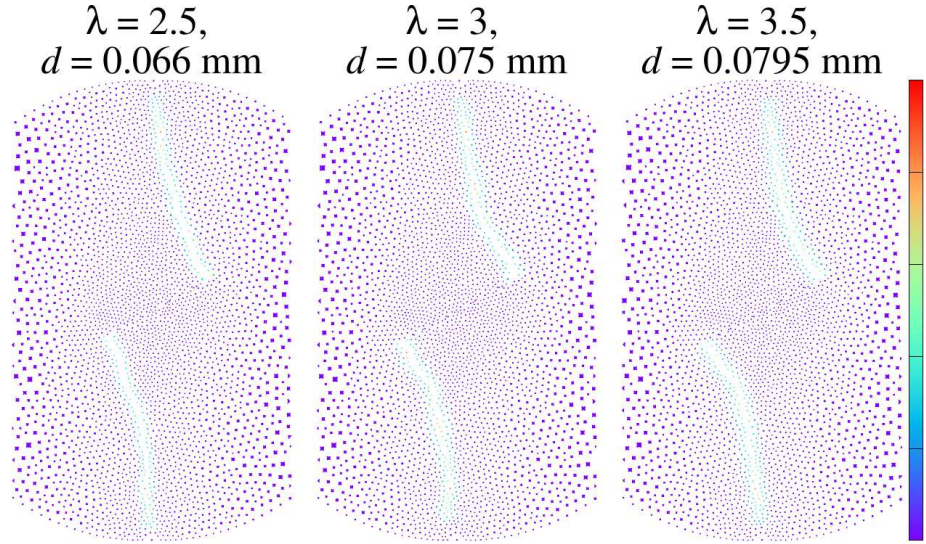}
	\caption{Disk tests with an inclined slot: damage degree and deformation plots (scale: 1:20) with $\alpha=60^\circ$}
	\label{fig:Disk1slot60deg}
\end{figure}

For disk tests with an inclined slot, the force-displacement curves and normalized peak loads are shown in Figure~\ref{fig:Disk1slotForces}, indicating agreeable results comparing to the results provided by phase field method \cite{Zhou2019} and Cracking Elements Method \cite{Yiming:20}.  The damage degree and deformation plots considering $\alpha=30^\circ$ and $\alpha=60^\circ$ are shown in Figures~\ref{fig:Disk1slot30deg} and~\ref{fig:Disk1slot60deg}.

\begin{figure}[htbp]
	\centering
	\includegraphics[width=0.85\textwidth]{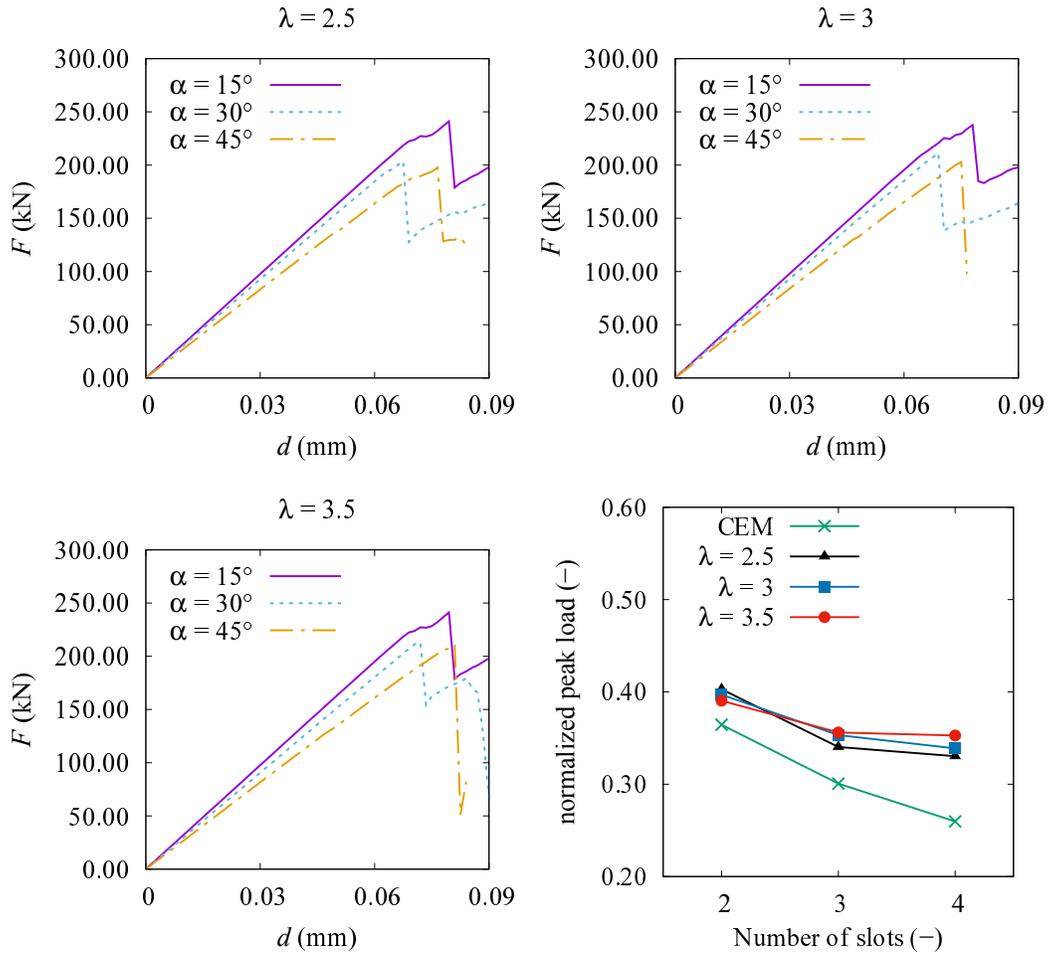}
	\caption{Disk tests with multiple slots: force-displacement curves and normalized peak loads comparing to the results obtained with phase field method provided with CEM in \cite{Yiming:21}}
	\label{fig:DiskMslotForces}
\end{figure}

\begin{figure}[htbp]
	\centering
	\includegraphics[width=0.85\textwidth]{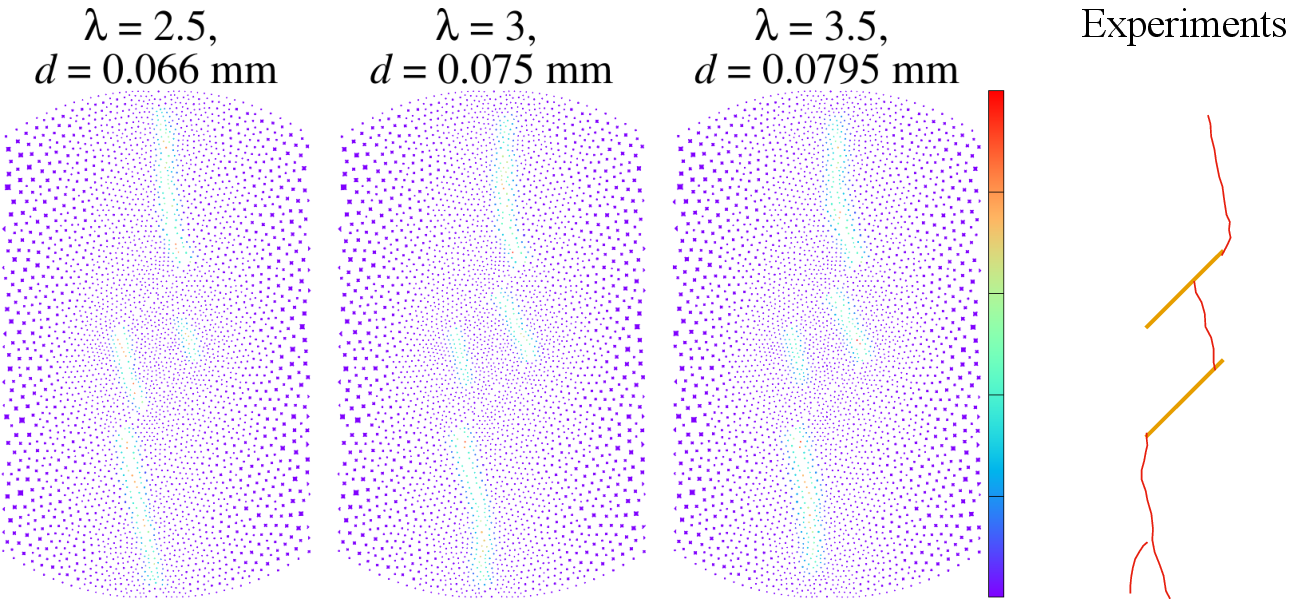}
	\caption{Disk tests with 2 slots: damage degree and deformation plots (scale: 1:20) comparing to the experimental results provided in \cite{Haeri2015}}
	\label{fig:DiskMslot2Re}
\end{figure}

\begin{figure}[htbp]
	\centering
	\includegraphics[width=0.85\textwidth]{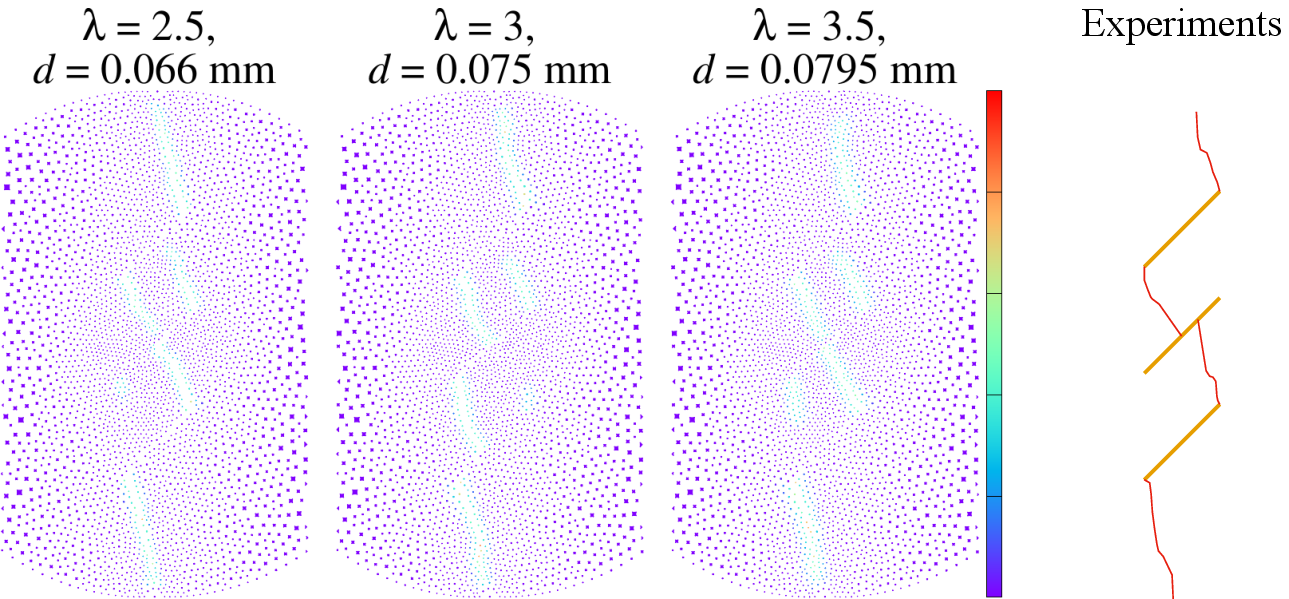}
	\caption{Disk tests with 3 slots: damage degree and deformation plots (scale: 1:20) comparing to the experimental results provided in \cite{Haeri2015}}
	\label{fig:DiskMslot3Re}
\end{figure}

\begin{figure}[htbp]
	\centering
	\includegraphics[width=0.85\textwidth]{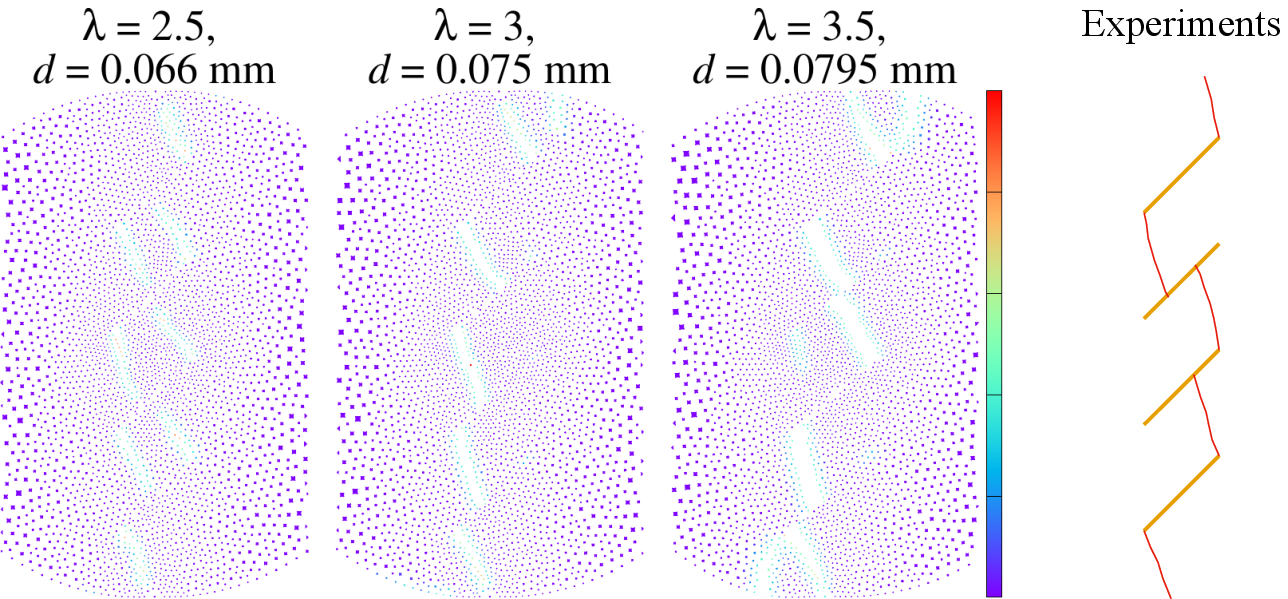}
	\caption{Disk tests with 4 slots: damage degree and deformation plots (scale: 1:20) comparing to the experimental results provided in \cite{Haeri2015}}
	\label{fig:DiskMslot4Re}
\end{figure}

For disk tests with multiple slots, the force-displacement curves and normalized peak loads are shown in Figure~\ref{fig:DiskMslotForces}, indicating weak dependency between the results and $\lambda$ after using adjusted $F_t$.  The damage degree and deformation plots are shown in Figures~\ref{fig:DiskMslot2Re} to~\ref{fig:DiskMslot4Re}, comparing to the experimental results provided in \cite{Haeri2015}.  Generally the patterns of cracking are similar.

\subsection{Plate with an inclined slot}
\begin{figure}[htbp]
	\centering
	\includegraphics[width=0.85\textwidth]{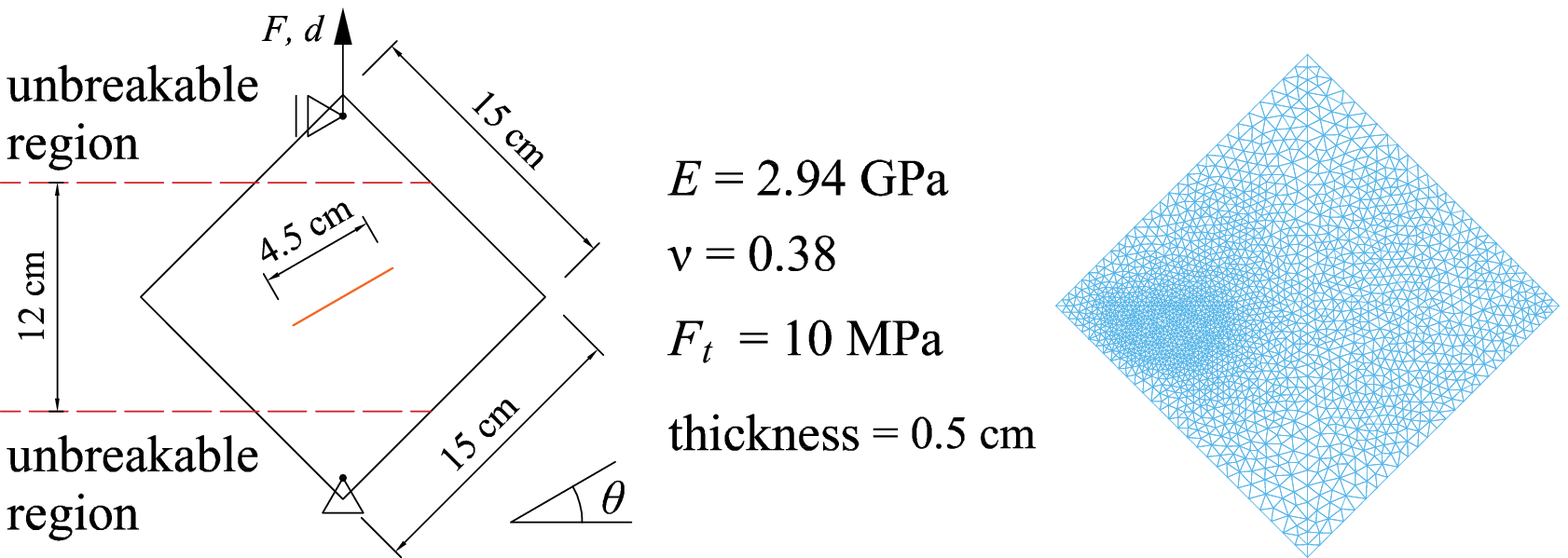}
	\caption{Plate with an inclined slot: model, material and mesh}
	\label{fig:PlateModel}
\end{figure}

\begin{figure}[htbp]
	\centering
	\includegraphics[width=0.7\textwidth]{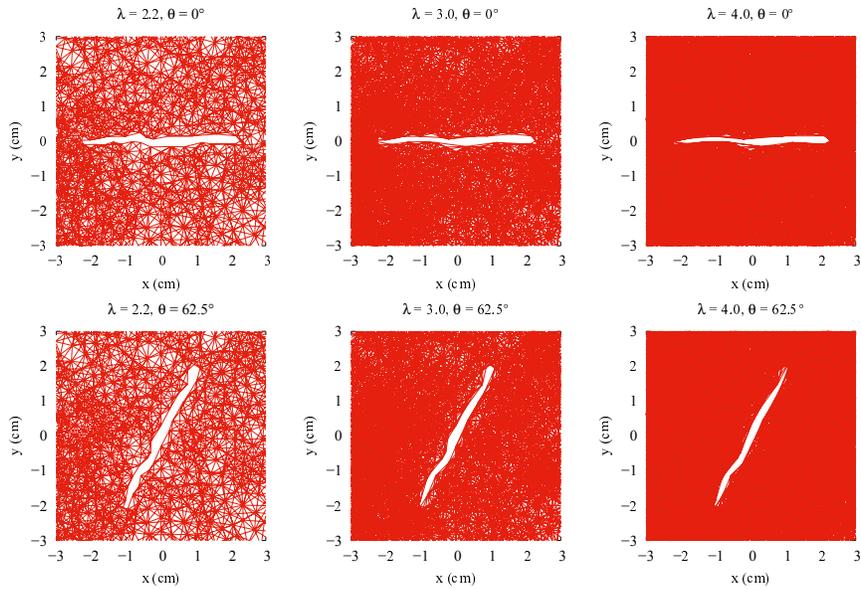}
	\caption{Plate with an inclined slot: bonds around the crack tips for cases $\theta=0^\circ$ and $\theta=62.5^\circ$}
	\label{fig:Platebonds}
\end{figure}

Plate made of PMMA with an inclined slot is a benchmark test for PD provided in \cite{Pdtheory}, which was experimentally investigated in \cite{AYATOLLAHI20091563}.  In the experiments, the cracks propagate axis-symmetrically.  The model, material and mesh of the test are shown in Figure~\ref{fig:PlateModel}.  The bonds intersect with the slot are removed for implicitly modeling the slot.  This strategy inevitably makes the crack tips a little coarse, see Figure~\ref{fig:Platebonds}.  This example is used for testing the influences of mesh on crack propagation.  There is a refined region on the left side of the model for checking whether the crack will be attracted by this region.  Different values of $\lambda$ are considered and the values of $F_t$ is adjusted based on Eq.~\ref{eq:Lft}, as: i) $\lambda=2.2, F_t=14.286$~MPa, ii) $\lambda=3, F_t=10$~MPa, and iii) $\lambda=4, F_t=7.273$~MPa.

\begin{figure}[htbp]
	\centering
	\includegraphics[width=0.55\textwidth]{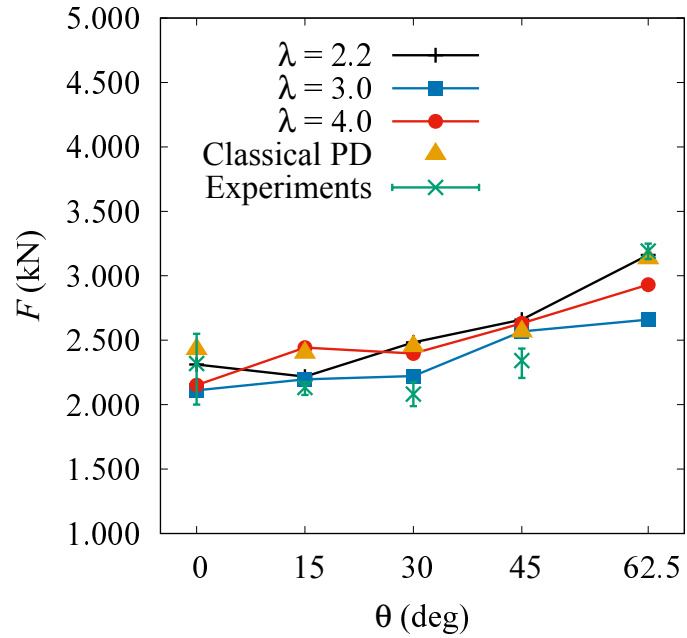}
	\caption{Plate with an inclined slot: the relationship between the peak loads and the inclined angle, comparing to the experimental results provided in \cite{AYATOLLAHI20091563} and numerical results by classical PD in \cite{Pdtheory}}
	\label{fig:PlateFs}
\end{figure}

\begin{figure}[htbp]
	\centering
	\includegraphics[width=0.9\textwidth]{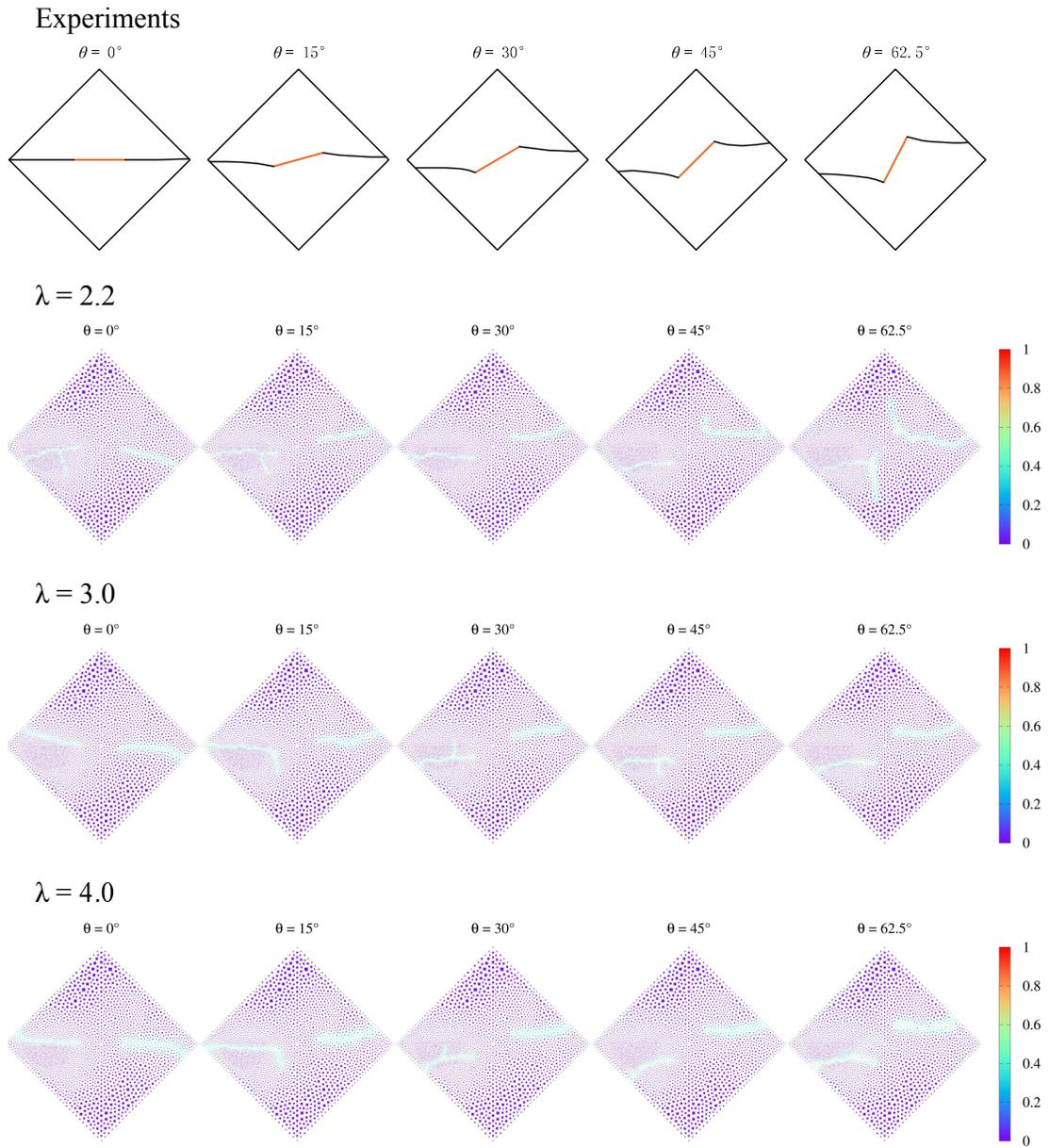}
	\caption{Plate with an inclined slot: the damage degree plots comparing to the experimental results provided in \cite{AYATOLLAHI20091563}}
	\label{fig:PlateCk}
\end{figure}

The relationship between the peak loads and the inclined angle is illustrated in Figure~\ref{fig:PlateFs}.  Considering different values of $\lambda$, after adjusting $F_t$, the obtained peak loads are generally similar.  The damage degree plots are shown in Figure~\ref{fig:PlateCk}.  In most cases, similar to the experiments, axis-symmetrical cracks (damaged regions) are obtained.  The crack is not attracted by the refined region on the left side.  Some unexpected branches are found in some cases, such as in the case with $\lambda=2$, $\theta=62.5^\circ$, which we attribute mainly to the coarse modeling of the crack tips.

\section{Conclusions}
\label{sec:conc}
In this work, we present a micropolar peridynamics model with non-unified horizon (NHPD).  The main features are summarized as
\begin{itemize}
	\item 
	In the pre-processing step, normal FEM discretization is used for providing material points.  The horizon varies with different points the size of which depends on the shortest distance between neighboring points.  The ratio of size of horizon to the shortest distance equals to a prescribed value $\lambda$.  When $\lambda$ increases, the non-local effects are enhanced;
	\item
	An iterative domain correction strategy is proposed for assuring the equivalence of strain energy density.  Then, based on the maximum strain energy density, a novel failure criterion is proposed for the NHPD which bridges the critical stretch to the mechanical strength $F_t$;
	\item
	Considering numerical studies regarding different values of $\lambda$ and different meshes, the results indicate NHPD shows generally weak mesh dependency.  Moreover, it is found that if $\lambda \ge 2$, $\lambda$ has weak influences on the stiffness of the structure while $\lambda$ has great influence on the equivalent strength of the structure $F^{eq}_t$.  A linear relationship between $F^{eq}_t\ /\ F_t$ and $\lambda$ is obtained and $F^{eq}_t=F_t$ when $\lambda=3$;
	\item
	Considering the linear relationship between $F^{eq}_t\ /\ F_t$ and $\lambda$ then adjusting the inputed $F_t$, similar results can be obtained regarding different values of $\lambda$.
\end{itemize}
The NHPD shows another routine for developing peridynamics models and the relationship between equivalent strength and $\lambda$ indicates correlations between strength and local/non-local damages.

\section{Acknowledgement}
The authors gratefully acknowledge financial support by the National Natural Science Foundation of China (NSFC) (51809069) and by the Hebei Province Natural Science Fund E2019202441 and the 2019 Foreign Experts Plan of Hebei Province.




\clearpage
\bibliographystyle{ieeetr}
\bibliography{Reference}







\end{document}